\documentclass{PoS}

\usepackage{amsmath}

\title{Baryon resonances and hadronic interactions in a finite volume}

\ShortTitle{Baryon resonances in a finite volume}

\author{\speaker{J. M. M. Hall}, A. C. P. Hsu, D. B. Leinweber, 
A. W. Thomas and R. D. Young  
\\
        Special Research Centre for the Subatomic Structure of
  Matter (CSSM), School of Chemistry and Physics, 
University of Adelaide 5005, Australia\\
        E-mail: 
\email{jonathan.hall@adelaide.edu.au
}}

\abstract{
In a finite volume, resonances and multi-hadron states 
are identified by discrete energy 
levels. When comparing the results of lattice QCD calculations
 to scattering experiments, 
it is important to have a way of associating the energy 
spectrum of the finite-volume 
lattice with the asymptotic behaviour of the S-matrix. 
A new technique for comparing energy eigenvalues 
with scattering phase shifts is introduced, which involves 
the construction of an exactly solvable matrix Hamiltonian model. 
The model framework is applied to the case of $\Delta\rightarrow N\pi$ decay, 
but is easily generalized to include multi-channel scattering. 
Extracting resonance parameters  
involves matching the energy spectrum of the model 
to that of a lattice QCD calculation. 
The resulting fit parameters are then used 
to generate phase shifts.  
Using a sample set of pseudodata, 
it is found that the extraction of the resonance 
position is stable with respect to volume for a variety 
of regularization schemes, and 
compares favorably with the well-known L\"{u}scher method. 
The model-dependence of the result 
is briefly investigated.  
}

\FullConference{The 30 International Symposium on 
Lattice Field Theory - Lattice 2012,\\
		June 24-29, 2012\\
		Cairns, Australia}

\begin{document}

\section{Introduction}

L\"{u}scher's method \cite{Luscher:1990ux} 
constitutes the principal method for relating 
the discrete energy spectrum calculated in lattice QCD 
with the continuous asymptotic states measured in hadron 
scattering experiments \cite{He:2005ey,Bernard:2008ax,Bernard:2009mw,Doring:2011nd,Doring:2011vk,Giudice:2012tg,Kreuzer:2012sr,Gockeler:2012yj}. 
By matching the asymptotic behaviour of the $S$-matrix 
onto the energy spectrum of the toroidal topology of the lattice, a 
geometric equation is obtained,
which is valid only for the scattering of two 
well-separated particles in a finite volume. 
In extending to more complicated cases 
 e.g. multi-channel scattering, the interpretation of L\"{u}scher's method 
becomes more difficult; however, such a development remains a promising 
and challenging area 
of ongoing research \cite{He:2005ey,Liu:2005kr,Bernard:2012bi,Doring:2012eu,Hansen:2012tf}. 

An alternative method for identifying resonance parameters in finite-volume 
scattering is proposed, which has the compelling property of being 
easily generalized to include more complicated interactions and additional 
channels. 
The method involves the construction of a matrix Hamiltonian model 
in a finite volume, such that 
its eigenvalue equation matches directly onto chiral effective field theory 
in the low-energy limit. 

As a test example,  the matrix Hamiltonian approach is applied to 
$\Delta\rightarrow N\pi$ decay. An energy spectrum can be generated from 
the model, and these energy levels can be matched directly 
to those of a lattice QCD 
calculation, fitting the free parameters of the model. With these 
fit parameters, 
the position of the resonance pole may then be obtained from the standard 
methods of scattering theory. 
The robustness of this Hamiltonian technique is tested by generating 
a finite-volume energy spectrum as `pseudodata', 
and matching this spectrum 
 to that of an alternative version of the model. Thus, a picture of the 
model-dependence is developed.

\section{The finite-volume matrix Hamiltonian model}

Consider a matrix Hamiltonian model for the 
$\Delta N\pi$ interaction, such that the finite-volume 
energy spectrum can be solved exactly. 
%
The Hamiltonian may be written as separate free and interaction parts,  
$H = H_0 + H_I$, where the free Hamiltonian includes the energies 
of the pion-nucleon system  
\begin{equation}
\label{H0}
 H_0  = 
\begin{pmatrix}
\Delta_0& 0 & 0 &\cdots\\
0 & \omega_\pi(k_1)& 0 &\cdots\\
0 & 0 &\omega_\pi(k_2)\\
\vdots &\vdots & & \ddots     \end{pmatrix},
\end{equation}
where $\omega_\pi(k_n) = \sqrt{k_n^2 + m_\pi^2}$, and 
$\Delta_0$ is the bare mass of the $\Delta$ baryon. The rows and columns 
of $H$ represent the momentum states of the pion relative to the nucleon. 
The values of momentum-squared available in a finite volume, $L^3$, are 
$k_n^2 = (\frac{2\pi}{L})^2(n_x^2 + n_y^2 + n_z^2)\equiv(\frac{2\pi}{L})^2n$, 
where $n$ is a squared integer. 

Including a direct coupling to a $\Delta$ baryon, 
the interaction 
Hamiltonian takes the form 
\begin{equation}
\label{HI}
 H_I  = \begin{pmatrix}
0& g^{\mathrm{fin}}_{\Delta N}(k_1) & g^{\mathrm{fin}}_{\Delta N}(k_2) &\cdots\\
  g^{\mathrm{fin}}_{\Delta N}(k_1)  & 0 &0 &\cdots\\
 g^{\mathrm{fin}}_{\Delta N}(k_2)  & 0 &0 &\cdots\\
\vdots & \vdots& \vdots&\ddots     \end{pmatrix}.
\end{equation}
The coupling, $g^{\mathrm{fin}}_{\Delta N}(k_n)$, is obtained from chiral effective field 
theory, and includes appropriate dimensional factors for a finite-volume 
calculation 
\begin{align}
g^{\mathrm{fin}}_{\Delta N}(k_n)  &= 
\sqrt{\frac{C_3(n)}{4\pi}}\left(\frac{2\pi}{L}\right)^{3/2}
g_{\Delta N}(k_n)\\
 &= 
\sqrt{\chi_\Delta\frac{C_3(n)}{2\pi^2}}\left(\frac{2\pi}{L}\right)^{3/2}
 \frac{k_n\, u(k_n)}{\sqrt{\omega_\pi(k_n)}},\\
\mbox{for}\quad
\chi_\Delta &= \frac{3}{32\pi f_\pi^2}\frac{2}{9}\,\mathcal{C}^2, 
\end{align}
using $f_\pi=92.4$ MeV, and the SU$(6)$ value $\mathcal{C}=-1.52$. 
The normalization 
$C_3(n)$ represents the number of ways of summing 
three squared integers to equal $n$. 
The introduction of a regulator function, $u(k_n)$, into the model 
 serves to keep the range of the interaction finite. In generating 
a sample set of pseudodata, 
 a dipole regulator with a 
mass of $\Lambda = 0.8$ GeV is chosen, being well-matched to phenomenology 
in the nucleon-pion sector \cite{Leinweber:1998ej,Leinweber:2005xz,Young:2009zb}. However, it 
will be demonstrated that the regularization scheme has limited impact 
on the consistency of the final 
extraction of the resonance position. 

The eigenvalue equation of the Hamiltonian, 
$\mathrm{det}(H-\lambda\mathbb{I})=0$, takes the form
\begin{align}
\label{eve1}
\lambda  &= \Delta_0 - \sum_{n=1}^{\infty}
\frac{\left(g^{\mathrm{fin}}_{\Delta N}(k_n) \right)^2}
{\omega_\pi(k_n)\, - 
\lambda}\\
\label{eve2}
&= \Delta_0 - 
\frac{\chi_\Delta}{2\pi^2}\left(\frac{2\pi}{L}\right)^3
\sum_{n=1}^{\infty}C_3(n)
\frac{k_n^2 u^2(k_n)}
{\omega_\pi(k_n)[\omega_\pi(k_n)\, - 
 \lambda]},
\end{align}
and the lowest-lying energy levels from the discrete spectrum of 
eigenvalues, $\lambda = E_j$, are shown as a function of lattice box size in 
Fig.~1. 
Note that the formulae in Eqs.~(\ref{eve1}) \& (\ref{eve2}) match the 
  one-loop $N\pi$ contribution to the $\Delta$ baryon self-energy 
 in effective field theory 
near the pole position: $\lambda\simeq E_\Delta\equiv 292$ MeV. 

At infinite volume, the real part of the one-pion loop integral 
takes the following form
\begin{align}
\mathrm{Re}\,\Sigma_{\Delta N}(k) &= \mathcal{P}\!\int_0^\infty\!\!\mathrm{d} 
k'\frac{k'\,^2 g_{\Delta N}^2(k'\, )}
{\omega_\pi({k} ) - \omega_\pi(k'\, )} \\
&= \chi_\Delta \frac{2}{\pi}\mathcal{P}\!\int_0^\infty\!\!\mathrm{d} 
k'\frac{k'\,^4\, u^2(k'\, )}
{\omega_\pi(k'\, )\,[\omega_\pi({k} ) - \omega_\pi(k'\, )]},
\end{align}
where $\mathcal{P}$ indicates that a principal value integral must  
be performed. This integral contributes to the phase shift 
via the $t$-matrix  for elastic 
$N\pi$ scattering (with a $\Delta$ baryon intermediate). 
The relationship between the phase shift and 
the on-shell quantity, $T=t(k,k;E^+)$, is
\begin{equation}
\label{tmat}
T = \frac{g^2_{\Delta N}({k} )}{\omega_\pi({k} ) - \Delta_0 
- \Sigma_{\Delta N}(k)} = 
-\frac{1}{\pi k \,\omega_\pi(k)} \,e^{i \delta(k)} \sin \delta(k).
\end{equation}
By solving Eq.~(\ref{tmat}), the phase shift, $\delta$, may be plotted as a 
function of energy, $E = \omega_\pi({k})$, to obtain the  
 curve  shown in Fig.~2. 
The bare resonance mass, $\Delta_0$, may be 
tuned so that the final resonance energy 
matches the physical $\Delta N$ mass-splitting, $E_\Delta \equiv 292$ MeV, 
and hence serves as an input for solving the finite-volume 
spectrum, via Eq.~(\ref{H0}).

\begin{figure}[t]
\begin{center}
\includegraphics[height=0.55\hsize,angle=90]{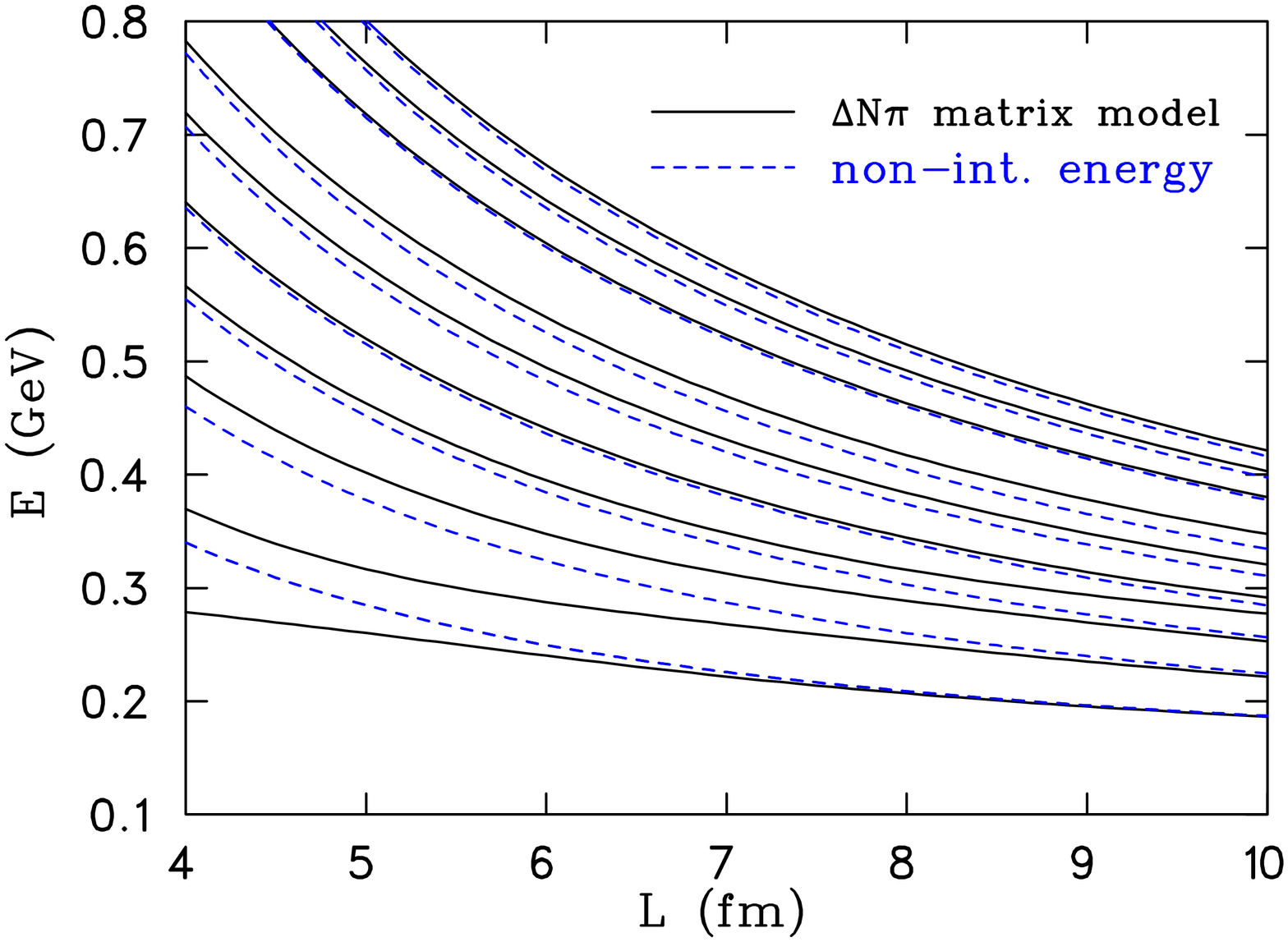}
\vspace{-11pt}
\caption{(color online). The lowest-lying energy levels from the $\Delta N \pi$ model 
(solid lines), and the corresponding non-interacting energies (dotted lines).}
\end{center}
\label{fig:EvsLfina}
%
%
\begin{center}
\includegraphics[height=0.550\hsize,angle=90]{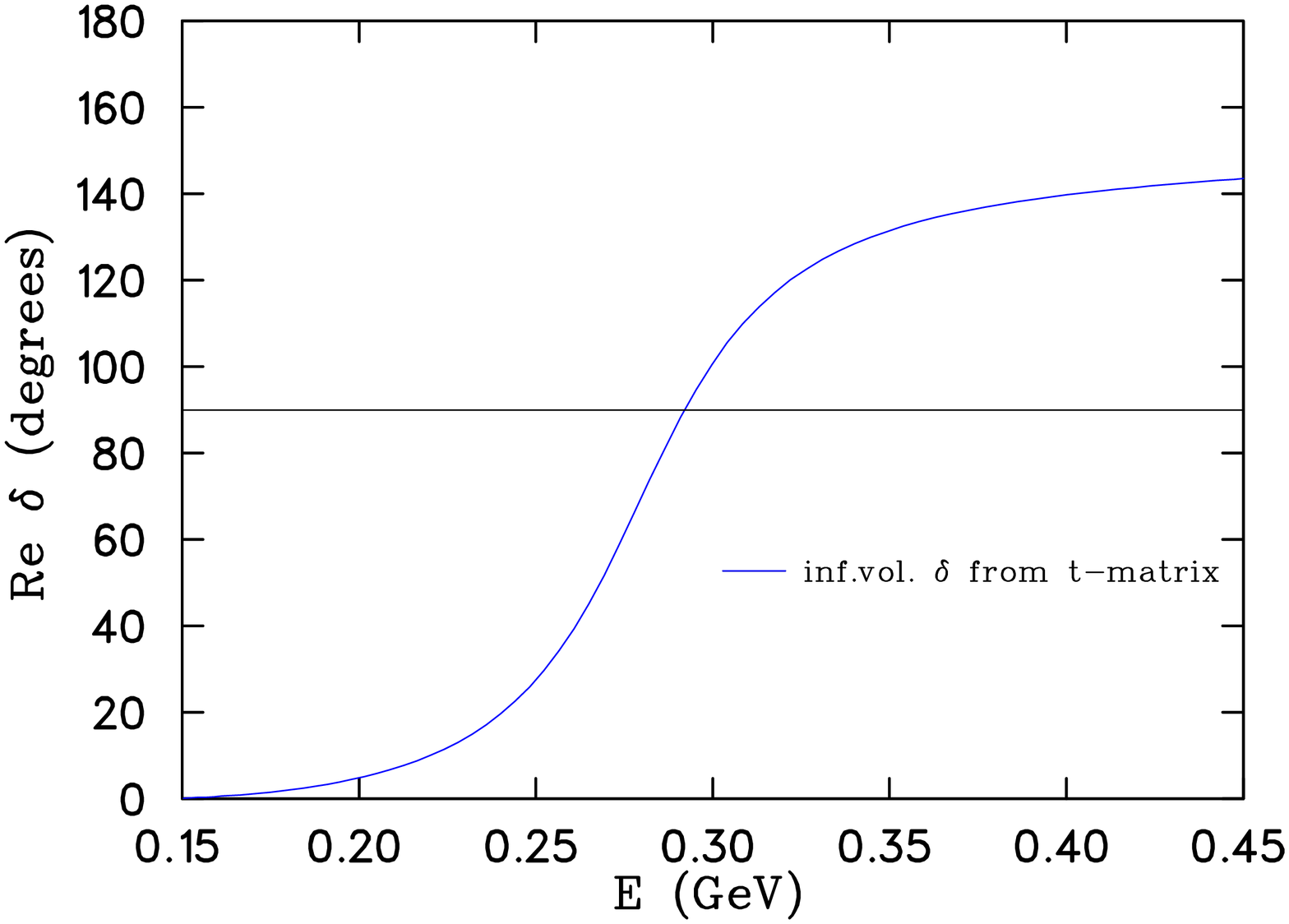}
\vspace{-11pt}
\caption{\footnotesize{(color online). The infinite-volume phase shift, $\delta$, associated with elastic $N\pi$ scattering via a $\Delta$ baryon intermediate state, plotted against the external pion energy, $E$ (where $E_\mathrm{tot} = M_N + E$), as calculated from the on-shell $t$-matrix.}}
\label{fig:dvsEinf}
\end{center}
\end{figure}

\section{L\"{u}scher's Method}

L\"{u}scher's formula describes a fixed relationship between the 
scattering phase shift, $\delta$, and the energy levels in a 
finite volume
\begin{equation}
\label{lf}
\delta(k_j;L) = j\,\pi - \phi\left(\frac{k_j L}{2\pi}\right),
\end{equation}
where $j$ is an integer indexing the energy levels, $E_j = \sqrt{k_j^2 + m^2}$.
L\"{u}scher's formula is derived assuming that two identical particles of mass 
$m$ 
scatter from a finite-range interaction (potential $V(r)=0$ for range $R<r$),  
and are well-separated ($R<r<L/2$), i.e. the particle wavefunction is 
in the asymptotic region. 
The angle function, $\phi(q)$, takes the form of a three-dimensional Zeta-like 
function (which must be regularized), 
defined in terms of dimensionless lattice momenta $q \equiv k L/(2\pi)$ 
\begin{align}
\label{phi}
\phi(q)&=-\mathrm{arctan}\left(\frac{\pi^{3/2} q}{\mathcal{Z}_{00}(1,q^2)}\right),
\\
\mathcal{Z}_{00}(1,q^2)&= \frac{1}{\sqrt{4\pi}}\sum_{\vec{n}\in\mathbb{Z}^3}
\frac{1}{\vec{n}^2-q^2}.
\end{align}
The momenta corresponding to a lattice QCD spectrum,  
 $k_j = \sqrt{E_j^2-m_\pi^2}$,  
may be input into L\"{u}scher's formula  
to obtain phase shifts, $\delta(k_j;L)$. 
Alternatively, the energy levels 
may be fit directly to the effective hadronic model described above, and the  
phase shift (and resonance position) 
 can then be extracted from the $t$-matrix in Eq.~(\ref{tmat}). 
Using a set of pseudodata generated from the $\Delta N\pi$ model, 
the two methods will be compared 
in the following section.

\section{An alternative method of phase shift extraction}

Motivated by the general result that the potential 
is separable near a resonance \cite{Lovelace:1964mq},   
a new method is proposed for obtaining a phase shift from discrete
 energy levels, such as those of lattice QCD. 
Using the $\Delta N\pi$ model, the parameters $\chi_\Delta$ and $\Delta_0$, 
 are chosen to minimize the chi-square between the energy levels of the model 
 and the energy levels of a lattice QCD calculation. 
An estimate of the phase shift can then be calculated from the 
$t$-matrix formula of Eq.~(\ref{tmat}),  
using the fitted values of the parameters. 

In order to test this idea, energy levels for the $\Delta N\pi$ model 
are treated as pseudodata. 
A modified version of the 
model is then constructed using a different regulator function, 
$\tilde{u}(k_n)$,
 such as a Gaussian regulator. 
By matching the two sets of energy levels 
and obtaining fit values for $\chi_\Delta$ and $\Delta_0$,  
phase shift estimates 
may be calculated for a range of box sizes, $L$. 

In order to visualize the comparison between L\"{u}scher's method and the
 new method, the  behaviour of the resonance energy, $E_{\mathrm{res}}$, may 
be plotted 
as a function of $1/L$, as shown in Fig.~3. 
Using L\"{u}scher's method, an interpolation function must be chosen 
in order to obtain the pole position from the phase shift. 
In the alternative method, 
two energy eigenvalues are chosen from the pseudodata, which are closest to 
the resonance energy, as estimated by L\"{u}scher's formula. 
These eigenvalues are then used to constrain the parameters 
$\chi_\Delta$ and $\Delta_0$.
Evidently, 
matching the pseudodata to a model with a different regulator function 
leads to a result that is at least comparable with L\"{u}scher's method.

By varying the regularization scale, $\Lambda$, a systematic uncertainty
 of only a few MeV is observed. 
This is encouraging, because it suggests that approximating the underlying 
physics of a lattice calculation with a regulator, sensibly chosen, 
will lead to a result that is not highly dependent on the features of 
the particular regulator function. 
Furthermore, one may treat $\Lambda$ as an additional fit parameter, 
and in this case, the closest three eigenvalues from the pseudodata
 are chosen for fitting. Fig.~3 indicates that a Gaussian regulator 
parameter of $\Lambda \simeq 0.6$ GeV provides the best matching 
with the pseudodata. 

The result of using the $\Delta N\pi$ model with regularization scale 
removed (i.e. $\Lambda\rightarrow\infty$) is also displayed in 
 Fig.~3. The pole extraction closely resembles that of L\"{u}scher's method. 

\begin{figure}[t]
\begin{center}
\includegraphics[height=0.6\hsize,angle=90]{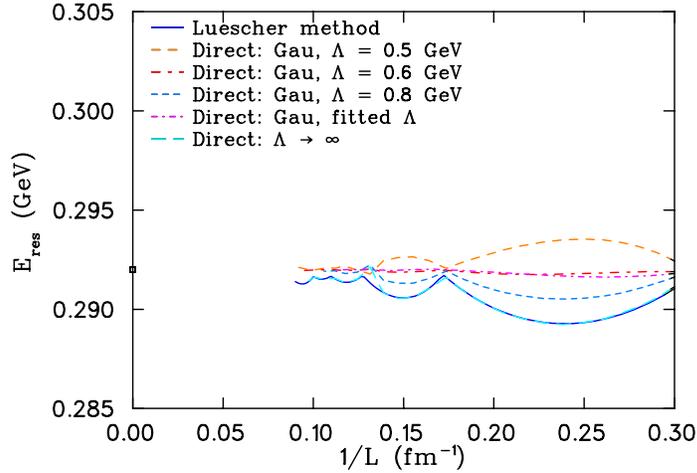}
\vspace{-11pt}
\caption{(color online). The resonance energy, $E_{\mathrm{res}}$, plotted against $1/L$. 
The experimental value is marked with a square. The results from fitting pseudodata with a different regulator, a Gaussian with $\Lambda = 0.5,\,0.6$ and $0.8$ GeV, or treating $\Lambda$ as a fit parameter, are plotted. The result of effectively removing the regulator (i.e. $\Lambda\rightarrow\infty$) is also shown. For comparison, the approach using L\"{u}scher's method is marked with a solid line.} 
\end{center}
\label{fig:EresvsLinv}
\end{figure}

%
%

\section{Summary}

An alternative
 method for the extraction of resonance parameters in a finite volume 
is investigated. An exactly solvable matrix Hamiltonian is constructed 
to model $\Delta\rightarrow N\pi$ decay in a finite-volume, 
 in anticipation of generalizing to 
more complicated multi-channel scattering problems. By matching the 
energy levels of the model to those of a lattice QCD calculation, the  
parameters of the model can be input into effective field theory 
in order to generate phase shifts. 
The model is tested by generating pseudodata, and extracting the 
resonance position using an 
alternative form of the model. 
The results are comparable with L\"{u}scher's method,  
and the extraction of the phase shift is stable with respect to volume. 

\section{Acknowledgements}
\vspace{-1mm}

We would like to thank M. D\"{o}ring and J. Dudek for helpful 
conversations. 
This research 
was supported by the Australian Research Council through 
 the ARC Centre of Excellence for Particle Physics at the 
Terascale, and through grants DP110101265 (DBL \& RDY), 
 FT120100821 (RDY) and FL0992247 (AWT).

\bibliographystyle{unsrt}
\bibliography{latticeref}

\end{document}